\documentclass{PoS}
\DeclareSymbolFont{newfont}{OML}{cmm}{m}{it}% Computer Modern math font
\DeclareMathSymbol{\Epsilon}{3}{newfont}{15}% Symbol 15
\usepackage[noadjust]{cite}

\title{\preprint{Constraining energy loss from high-$p_{\rm T}$ azimuthal asymmetries}}
\ShortTitle{High-$p_T$ asymmetries and initial stages}

\author{\speaker{Carlota Andres}$^a$ N\'estor Armesto$^{b}$, Harri Niemi$^{cd}$, Risto Paatelainen$^{de}$ and Carlos A. Salgado$^{b}$\\
\llap{$^a$} Jefferson Lab, Virginia, USA\\
\llap{$^b$} Instituto Galego de F\'isica de Altas Enerx\'ias, Universidade de Santiago de Compostela, Galicia, Spain\\ 
\llap{$^c$} University of Jyv\"askyl\"a, Finland\\
\llap{$^d$} Helsinki Institute of Physics, Finland\\
\llap{$e$} CERN, Geneva, Switzerland\\
E-mail: \email{carlota@jlab.org}, \email{nestor.armesto@usc.es}, \email{harri.m.niemi@jyu.fi}, \email{risto.sakari.paatelainen@cern.ch}, \email{carlos.salgado@usc.es}}

\abstract{The nuclear modification factor $R_{\rm AA}$ has been satisfactorily described by various jet quenching models. Nonetheless, all these formalisms, until very recently, underpredicted the high-$p_{\rm T}$ (> 10 GeV) elliptic flow $v_2$. We find that the simultaneous description of these observables requires to strongly suppress the quenching for the first $\sim 0.6$ fm after the collision. This shows the potential of jet quenching observables to constrain the dynamics of the initial stages of the evolution.}

\FullConference{International Conference on Hard and Electromagnetic Probes of High-Energy Nuclear Collisions\\
		30 September - 5 October 2018\\
		Aix-Les-Bains, Savoie, France}

\begin{document}
\section{Introduction}
Jet quenching phenomena arising from the modification of parton showers in a colored medium are considered one of the most successful and versatile tools to characterize the quark-gluon plasma (QGP) created in heavy-ion collisions at the Relativistic Heavy Ion Collider (RHIC) and at the Large Hadron Collider (LHC). While the nuclear modification factor $R_{\rm AA}$ has been properly described by all the energy loss models, their results on the high-$p_{\rm T}$ elliptic flow were systematically below the experimental data points. Recently, it was claimed that soft-hard correlations are essential to well describe the high-$p_{\rm T}$ asymmetries, whose correct definition is given by the scalar product, $v_{\rm n}^{\rm SP}$  \cite{Noronha-Hostler:2016eow}.

This talk is based on our recent publication \cite{Andres:2019eus}. We compute here the nuclear modification factor $R_{\rm AA}$ and the high-$p_{\rm T}$ elliptic flow $v_2$ for charged particle production in the 20 -- 30\% centrality class of $\sqrt{s_{\mathrm{NN}}}$ = 2.76 TeV Pb-Pb collisions at the LHC using the \textit{ Quenching Weights} (QWs) formalism \cite{Salgado:2003gb} and EKRT event-by-event (EbyE) hydrodynamic profiles \cite{Niemi:2015qia}. We show that the treatment of initial stages is decisive for the simultaneous description of both observables, since the high-$p_{\rm T}$ $v_2$ experimental data can only be described by turning off the quenching for the first $\sim 0.6$ fm after the collision.
\section{Framework}
The cross section of a hadron $h$ at rapidity $y$ and transverse momentum $p_{\rm T}$ can be written as
\begin{equation}
\frac{d\sigma^{AA\to h}}{dydp_{\rm T}} = \int dq_{\rm T}\,dz\frac{d\sigma^{AA\to k}}{dydq_{\rm T}}\,P(\Epsilon)\\,D_{k\to h}(z,\mu_{\rm F}\equiv p_{\rm T})\,\delta\left(p_{\rm T}-z(1-\Epsilon)q_{\rm T}\right).
\label{eq:crossec}
\end{equation}
All the calculations are performed at next-to-leading order (NLO) in perturbative Quantum Chromodynamics (pQCD) using the code from Ref.~\cite{Stratmann:2001pb}. For the parton distribution functions (PDFs) we make use of CTEQ6.6M \cite{Nadolsky:2008zw} proton PDFs with EPS09 \cite{Eskola:2009uj} nuclear corrections. We employ DSS07 \cite{deFlorian:2007aj} fragmentation functions. The Quenching Weights $P(\Epsilon)$, used in the multiple soft approximation \cite{Salgado:2003gb}, depend on two variables, $\omega_{\rm c}$ and $R$, which, for a dynamically expanding medium, are proportional, respectively, to the first and second moment of the jet quenching parameter $\hat{q}(\xi)$ given by:
\begin{equation} 
\hat{q} (\xi) \equiv \,K\cdot 2\,\varepsilon^{3/4}(\xi),
\label{eq:qhat}
\end{equation}
along the trajectory, $\xi$, of the radiating parton. Here $\varepsilon$ is the local energy density taken from EKRT hydrodynamic simulation. The $K$-factor is the only free parameter of our approach, which is fitted to the $p_{\rm T}$ > 10 GeV $R_{\rm AA}$ experimental data \cite{Abelev:2012hxa}. Then, the fitted $K$-factor is employed to compute the elliptic flow in the hard sector by means of the scalar product definition \cite{Noronha-Hostler:2016eow}.

Since Eq.~(\ref{eq:qhat}) is valid both for the partonic and hadronic phase of the evolution \cite{Baier:2002tc}, we can explore here two different scenarios: a) assuming no quenching in the hadronic phase, that is, stopping the energy loss at the chemical freeze-out $T_{\mathrm{q}} = T_{\mathrm{chem}}$ = 175 MeV or b) using Eq.~(\ref{eq:qhat}) in both phases, i.e., up to the kinetic freeze-out $T_{\mathrm{q}} = T_{\mathrm{dec}}$ = 100 MeV\footnote{Being $T_{\mathrm{q}}$ the temperature at which the energy loss ends.}.

We obtain the EbyE space-time distribution of the local energy density by solving the relativistic hydrodynamic equations with EKRT initial state, shear viscosity parametrization $\eta/s(T) = param1$  from Ref.\cite{Eskola:2009uj} and initialization proper time  $\tau_{\rm f} =$ 0.197 fm. Since energy loss in the BDMPS-Z formalism does not require neither thermalization nor isotropization, it can be employed for times prior to $\tau_{\rm f}$ and, thus, $\hat{q}(\xi)$ for these early times needs to be obtained via extrapolations. All the phenomenological studies aiming to determine the high-$p_{\rm T}$ harmonics start the interaction of the hard parton with the medium at the initial time of the hydrodynamic model, generally $\tau_{\rm f}$ = 0.6 fm (or larger). Since the initialization time of EKRT EbyE hydrodynamics  is $\tau_{\rm f}$ = 0.197 fm, we present a framework that, for the first time, allows us to change the quenching in the early stages of the evolution and, thus, to establish when the energy loss begins. For this purpose, we consider the three following early time extrapolations\footnote{Denoting by $\tau_{\rm q}$ the time where the jet quenching begins.}:\\
Case i) $\tau_{\rm q}=0$ fm, where $\hat{q}(\xi)=  \hat{q}(\tau_{\rm f})$ for $\xi < \tau_{\rm f} = 0.197$ fm.\\
Case ii) $\tau_{\rm q}=0.197$ fm. Here, $\hat{q}(\xi)=$ 0 for $\xi < \tau_{\rm f} = 0.197$ fm, that is, the energy loss begins at $0.197$ fm.\\
Case iii) $\tau_{\rm q}=0.572$ fm. Here, $\hat{q}(\xi)=$ 0 for $\xi < \tau_{\rm q} = 0.572$ fm. So, the quenching starts at $0.572$ fm.\\

\section{Results}
\begin{figure*}[ht]
\includegraphics[width=\textwidth,trim = {0 2cm 0 2cm}]{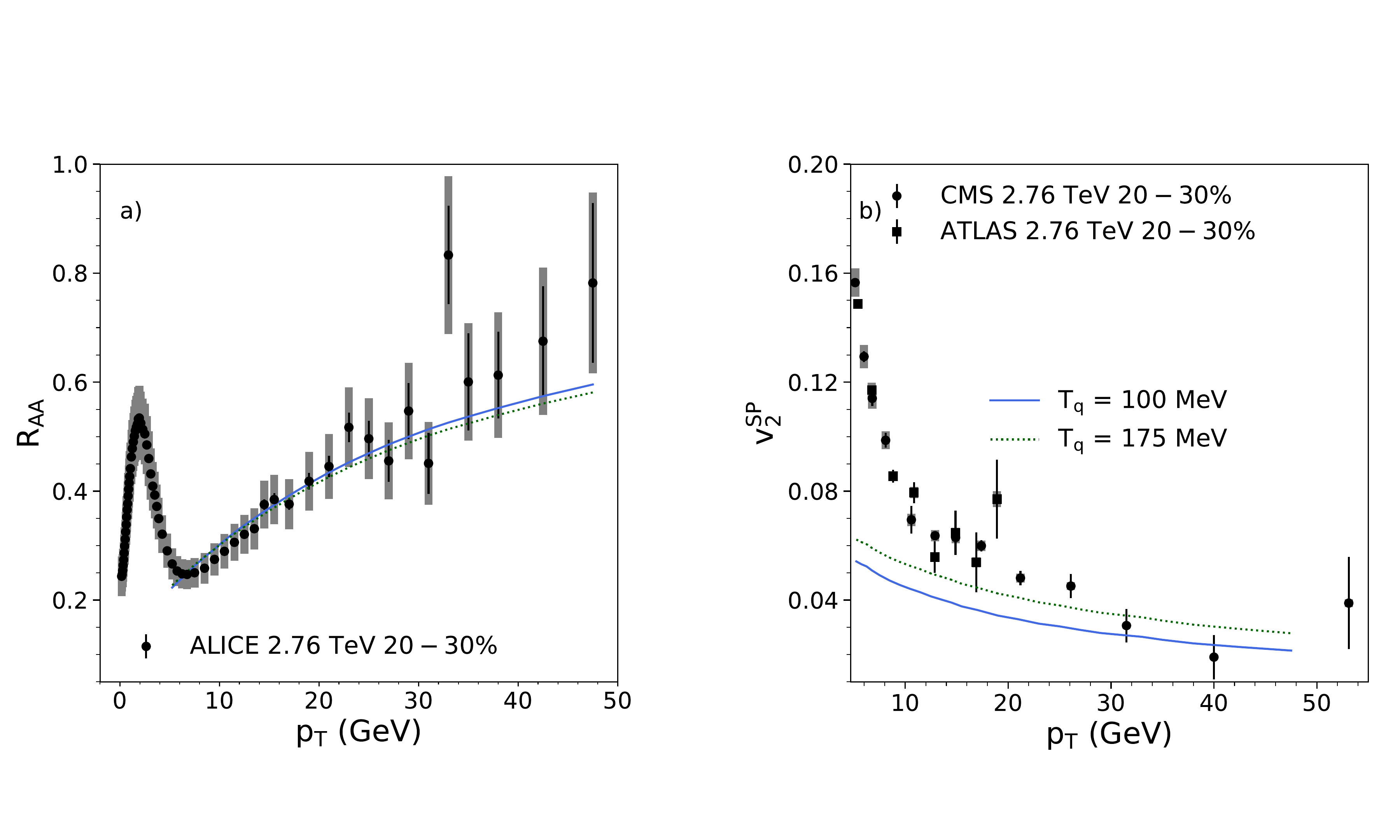}
\caption{(Color online) (a) $R_{\rm AA}(p_{\rm T})$ and (b) $v_2^{\rm SP}(p_{\rm T})$  for 20--30\% semi-central $\sqrt{s_{\mathrm{NN}}}$ = 2.76 TeV Pb-Pb collisions at the LHC compared to their respective experimental data  \cite{Abelev:2012hxa,Chatrchyan:2012xq,ATLAS:2011ah}. The blue solid line and green dotted lines correspond, respectively, to stopping the energy loss at the kinetic freeze-out $T_\mathrm{q} = T_{\mathrm{dec}}$ = 100 MeV and at the chemical freeze-out $T_\mathrm{q} = T_{\mathrm{chem}}$ = 175 MeV. Case ii) $\tau_{\rm q}=0.197$ fm is employed.}
\label{fig:Tq}
%\ref{fig:Tq}
\end{figure*}

The dependence of the nuclear modification factor $R_{\rm AA}(p_{\rm T})$ and the elliptic flow $v_2^{\rm SP}(p_{\rm T})$ in the hard sector on the end-point of the quenching is shown in Fig.~\ref{fig:Tq}. We assume here no energy loss before  the initial time $\tau_{\rm f}$ = 0.197 fm  of the hydrodynamic profile, that is, Case ii) $\tau_{\rm q}=0.197$. As it can be seen on the left panel of this figure, the dependence of the inclusive particle production suppression on the presence or absence of quenching in the hadronic phase is mild. The right panel of Fig.~\ref{fig:Tq} shows that the high-$p_{\rm T}$ $v_2^{\rm SP}(p_{\rm T})$ is sensitive to the end-point of the energy loss, suggesting a better description of the experimental measurements when there is only energy loss in the partonic phase. It is worth emphasizing that, independently of when we end our calculation, the high-$p_{\rm T}$ elliptic flow remains underestimated.

\begin{figure*}[ht]
\includegraphics[width=\textwidth,trim = {0 2cm 0 0.7cm}]{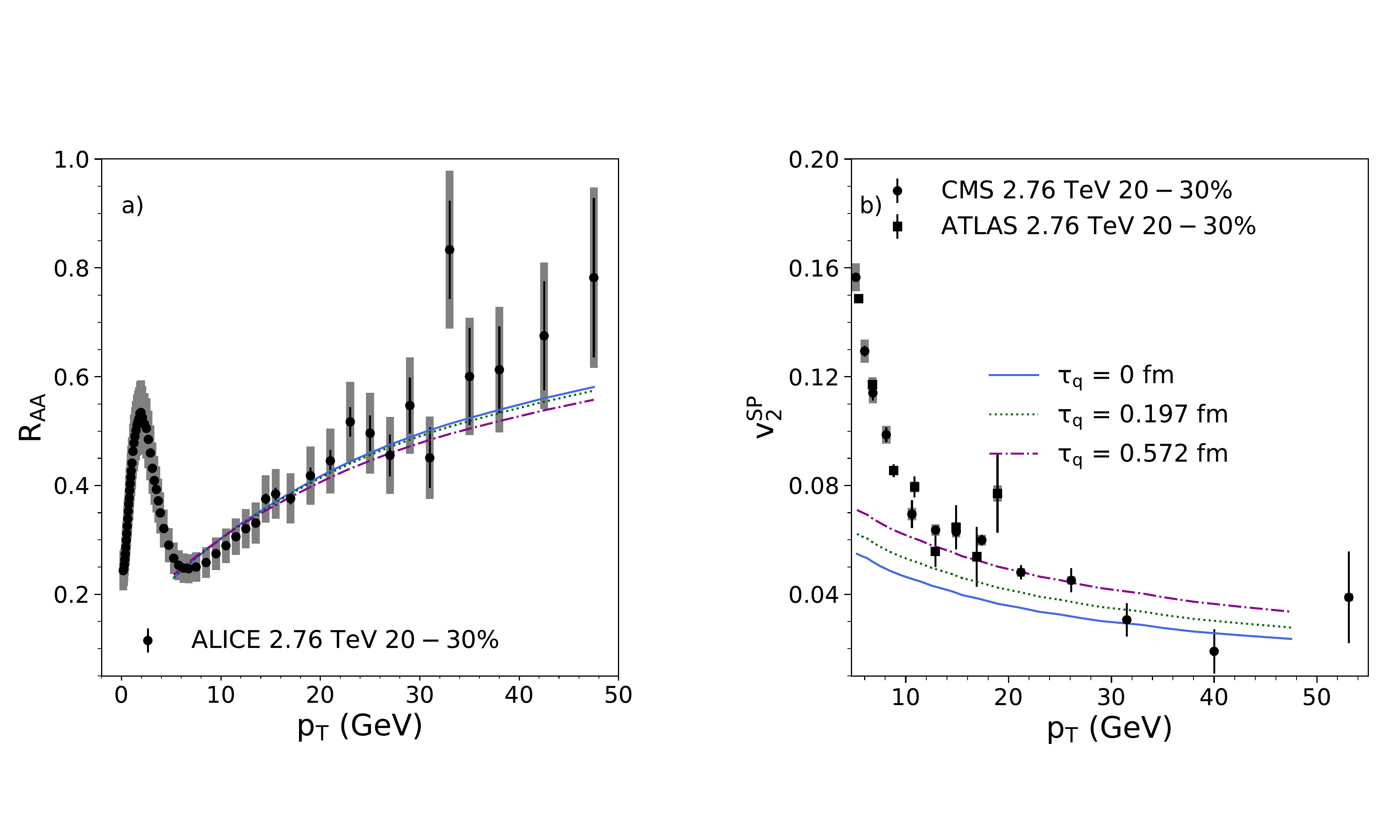} 
\caption{(Color online) (a) $R_{\rm AA}(p_{\rm T})$ and (b) $v_2^{\rm SP}(p_{\rm T})$ for the 20--30\% centrality class in $\sqrt{s_{\mathrm{NN}}}$ = 2.76 TeV Pb-Pb collisions at the LHC compared to their respective experimental data  \cite{Abelev:2012hxa,Chatrchyan:2012xq,ATLAS:2011ah}. The blue solid, $\tau_{\rm q}=0$ fm, dotted green, $\tau_{\rm q}=0.197$ fm, and dashed purple, $\tau_{\rm q}=0.572$ fm, lines correspond, respectively, to Cases i), ii) and iii) of the early times treatment. $T_{\mathrm{q}} = T_{\mathrm{chem}}$ = 175 MeV is used.}
\label{fig:tauq}
%\ref{fig:tauq}
\end{figure*}

In Fig.~\ref{fig:tauq} we present the dependence of these observables on the starting time of the energy loss. We assume here that there is no quenching in the hadronic phase, i.e., $T_{\mathrm{q}} = T_{\mathrm{chem}}$.
It can be seen that, no matter when the quenching begins, our model fairly describes the nuclear modification factor. Nevertheless, the high-$p_{\rm T}$ elliptic flow turns out to be very sensitive to the starting point of the energy loss. In fact, the correct determination of the $v_2^{\rm SP}(p_{\rm T})$ in the hard sector within our approach requires the absence of energy loss for the first $\sim 0.6$ fm after the collision, in agreement with the set-up that other works were implicitly employing \cite{Noronha-Hostler:2016eow,Betz:2016ayq,Zigic:2018ovr,Shi:2018lsf}.

\section{Conclusions} 
We compute the azimuthally averaged $R_{\rm AA}$ and the high-$p_{\rm T}$ $v_2$ for the 20 -- 30\% centrality class in $\sqrt{s_{\mathrm{NN}}}$ = 2.76 TeV Pb-Pb collisions at the LHC. Our framework is based on the QWs embedded in the EKRT EbyE hydrodynamic model. We study the sensitivity of these observables to the presence of energy loss during the hadronic phase of the evolution, and to the starting time of the energy loss. The simultaneous and correct determination of these observables within our approach shows up to be incompatible with the presence of energy loss during the first $\sim$0.6 fm after the collision. This corresponds to the set-up that, up to now, all the phenomenological studies trying to describe the jet harmonic coefficients were implicitly adopting. Therefore, this is not a special characteristic of our framework, but a general outcome: the azimuthal asymmetries in the hard sector emerge as a direct signature of the dynamics of the initial stages of the evolution.

\section*{Acknowledgements}
We acknowledge the CSC -- IT Center for Science in Espoo, Finland, for the allocation of the computational resources. HN is supported by the Academy of Finland, project 297058. RP is supported by the European Research Council, grant no. 725369. NA and CAS are supported by Ministerio de Ciencia e Innovaci\'on of Spain under project FPA2017-83814-P and Unidad de Excelencia Mar\'{\i}a de Maetzu under project MDM-2016-0692, by Xunta de Galicia (Conseller\'{\i}a de Educaci\'on) and FEDER. This work has been performed in the framework of COST Action CA15213 `Theory of hot matter and relativistic heavy-ion collisions' (THOR). 

\bibliography{mybibfile}

\end{document}